\documentclass[a4paper]{revtex4}

\usepackage[%
	final,%
]{graphicx}

\usepackage[
   centertags, % (default) center tags vertically
   sumlimits,  %(default) Place the subscripts and superscripts of summation
   intlimits,  % Like sumlimits, but for integral symbols.
   namelimits, % (default) Like sumlimits, but for certain 'operator names' such as
]{amsmath} %
\usepackage{amssymb}

\usepackage[%
	pdftitle={Third-order dissipative hydrodynamics from the entropy principle},%
	pdfauthor={Andrej El, Zhe Xu, and Carsten Greiner},%
	pdfsubject={Third-order dissipative hydrodynamics from the entropy principle},
	pdfstartview=FitH,
	pdfpagemode=UseNone,
	bookmarksopen=true
	]{hyperref}

\usepackage[all]{hypcap} % Links to figures point actually on figure itself, not on caption

\usepackage{dcolumn} % tabellenasurichtung am komma

\usepackage{multirow} % tabelleneinträge über mehrere Zeilen

\usepackage{microtype} % bessere trennung

\makeatletter
\def\tagform@#1{\maketag@@@{\ignorespaces#1\unskip\@@italiccorr}}
\let\orgtheequation\theequation
\def\theequation{(\orgtheequation)}
\makeatother

\newcommand{\beq}{\begin{equation}}
\newcommand{\eeq}{\end{equation}}

\begin{document}
\title{Third-order dissipative hydrodynamics from the entropy principle}

\author{Andrej El$^1$\footnote{el@th.physik.uni-frankfurt.de}, Zhe Xu$^{2,1}$, and Carsten
Greiner$^1$}

\address{$^1$ Institut f\"ur Theoretische Physik, Johann Wolfgang 
Goethe-Universit\"at Frankfurt, Max-von-Laue-Str. 1, 
D-60438 Frankfurt am Main, Germany}
\address{$^2$ Frankfurt Institute for Advanced Studies, Ruth-Moufang-Str. 1, D-60438 Frankfurt am
Main, Germany}

\begin{abstract}
We review the entropy based derivation of third-order hydrodynamic equations and compare their
solutions in one-dimensional boost-invariant geometry with calculations by the partonic cascade
BAMPS. We demonstrate that Grad's approximation, which underlies the derivation of both
Israel-Stewart and third-order equations, describes the transverse spectra from BAMPS with high
accuracy. At the same time solutions of third-order equations are much closer to BAMPS results than
solutions of Israel-Stewart equations. Introducing a resummation scheme for all higher-oder
corrections to one-dimensional hydrodynamic equation we demonstrate the importance of higher-order
terms if the Knudsen number is large.
\end{abstract}

\maketitle

A causal theory of relativistic dissipative hydrodynamics was first 
formulated by Israel and Stewart \cite{IS} and has been successfully 
applied to study a wide range of ultra-relativistic heavy-ion 
collision phenomena \cite{Heinz:2009xj,Romatschke:2009im,Teaney:2009qa}. However, recently
presented detailed comparisons of solutions of Israel-Stewart equations and kinetic transport
calculations have demonstrated that deviations between them increase with increasing strength of
dissipation. Israel-Stewart equations can be derived from the divergence of the off-equilibrium
entropy current, which is expanded up to second-order in dissipative fluxes. It is thus of interest
to investigate whether a better agreement between hydrodynamic and kinetic transport calculations
can be achieved if the entropy current is expanded one order higher than in Israel-Stewart theory.

The underlying equation of the kinetic transport theory is the Boltzmann equation
\beq
p^\mu \partial_\mu f(x,p) = C[f(x,p)]
\eeq
which describes the space-time evolution of the phase-space particle distribution
$f(x,p)=\frac{d N}{d^3p d^3 x}$ due to drift and diffusion on the left hand side and the collision
processes on the right hand side of the equation. A connection
between the microscopic kinetic transport theory and the macroscopic theory of hydrodynamics can be
established using the Grad's method, in which the off-equilibrium distribution is approximated by
\beq
f_{off-eq} = f_0 \left( 1 + \epsilon + \epsilon_\mu p^\mu + \epsilon_{\mu\nu} p^\mu p^\nu \right)
\label{foffeq}
\eeq 
where $f_0$ denotes the isotropic (equilibrium) distribution and the fields $\epsilon,
\epsilon_\mu$ and $\epsilon_{\mu\nu}$ are related to the dissipative fluxes $\Pi, q^\mu$ and
$\pi^{\mu\nu}$. The exact form of $\epsilon, \epsilon_\mu$ and $\epsilon_{\mu\nu}$ can be obtained
from the definitions of dissipative currents and the matching conditions
\beq
u_\mu u_\nu (T^{\mu\nu} - T_{eq}^{\mu\nu}) = 0 ,~~~ u_\nu (N^{\nu} - N_{eq}^{\nu}) = 0
\label{matching}
\eeq
where $T_{eq}^{\mu\nu}$ and $N_{eq}^{\nu}$ are the energy-momentum tensor and the particle current
in equilibrium. 
 
In the following we consider a massless gas of Boltzmann particles (gluons) undergoing
a boost-invariant one dimensional expansion \cite{Bjorken:1982qr}. For the considered system the
equation of state is
$e=3p$. Bulk pressure and heat flux vanish identically  and the local rest frame off-equilibrium
distribution becomes \cite{El09,El10}
\beq
f(x,p) = d_g e^{-E/T} \left( 1 + \frac{3}{8 e T^2} \pi_{\mu\nu} p^\mu p^\nu \right)
\label{foffeq_B}
\eeq
with the energy density $e$ and the degeneracy factor for gluons, $d_g=16$. The matching conditions
\ref{matching} allow to define the
temperature $T$ for an off-equilibrated system by matching its energy and particle densities $e$
and $n$ to a fictitious equilibrium state: 
\beq
e = e_{eq},~~~ n = n_{eq}
\eeq
For the Boltzmann gas considered here the temperature $T$ in Eq.\ref{foffeq_B} is then
\beq
T = \frac{e}{3n} .\
\eeq
Grad's approximation, Eqs.\ref{foffeq} resp. \ref{foffeq_B}, is essential for derivations of
hydrodynamic equations from the Boltzmann equation \cite{Denicol10, R09} resp. from the entropy
principle \cite{M04,El10}. It is thus important to quantify how accurate the
approximation in Eq.\ref{foffeq_B} can reproduce the off-equilibrium distribution obtained from the
numerical solution of the Boltzmann Equation in the partonic cascade BAMPS \cite{XG05}. BAMPS has
recently been applied to investigate a wide range of phenomena
such like the buildup of the elliptic flow \cite{XG09}, the energy loss
of high energy gluons \cite{F09}, the extraction of the second-order 
viscosity coefficient \cite{El09}, and the formation and propagation
of shock waves \cite{B09} in ultra-relativistic heavy-ion
collisions.

For the results presented in this work we use BAMPS calculations with elastic isotropic cross
section adjusted in such a way that the shear viscosity to entropy density ratio is constant
throughout the evolution, as introduced in \cite{HM09,El10}. The initial condition is a Boltzmann
distribution with $T_0=500 MeV$ at the initial time $\tau_0=0.4 fm/c$. All results are extracted
from the central rapidity bin $\eta \in \left[-0.1:0.1 \right]$.

To quantify the deviations of the expression in Eq.\ref{foffeq_B} from the actual distribution in
BAMPS we take the ratio $\delta^{BAMPS}_{Grad}$ of the transverse distributions calculated as
follows:
\beq
\delta^{BAMPS}_{Grad}=\frac{\left( dN/p_T/dp_T \right)_{BAMPS}}{\left( dN/p_T/dp_T
\right)_{Grad}} = \frac{\big< p_0 f_{BAMPS}  \big>_{y,\varphi}}{\bigg< p_0 d_g \lambda
e^{\frac{-p_T \cosh y}{T}} \bigg( 1 + \frac{3}{8
T^2}\frac{\pi}{e}p_T^2\big(\frac{1}{2}-\sinh^2 y\big) \bigg)
\bigg>_{y,\varphi}} .\
\label{delta_grad}
\eeq
In the latter expression $\pi$ denotes the shear pressure, which is the dissipative correction to
the longitudinal pressure: 
\beq
\pi = -\pi^{33} = T_{eq}^{33} - T^{33}
\eeq
\begin{figure}
\includegraphics[width=7.0cm,angle=-90]{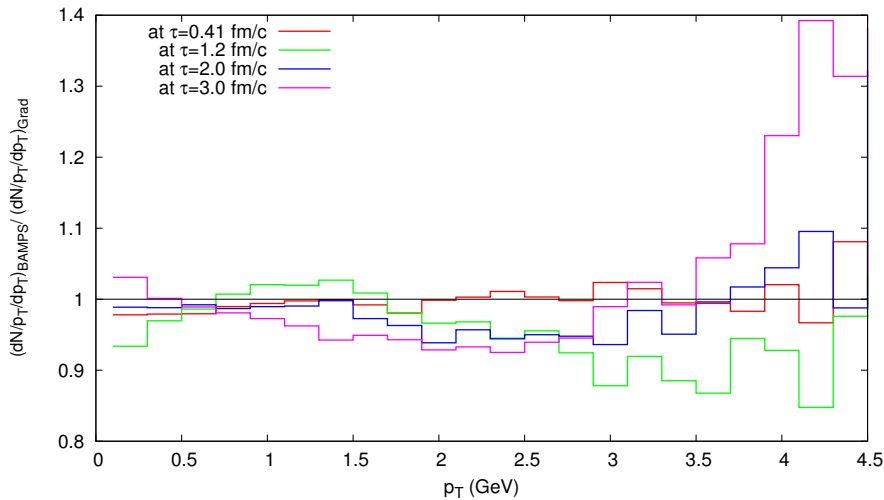}
\caption{Ratio of transverse particle distribution from BAMPS with $\eta/s=0.4$ to the one
calculated by Eq.\ref{foffeq_B} at different times with $e$, $T$, $\pi^{\mu\nu}$ extracted from
BAMPS.}
\label{fig:delta_BG}
\end{figure}
For the analytic calculation using Eq.\ref{foffeq_B} the values of $e$ and $T$ as well as the
components of $\pi_{\mu\nu} = T_{\mu\nu} - T^{eq}_{\mu\nu}$ are extracted from BAMPS. The deviation
$\delta^{BAMPS}_{Grad}$ is shown in Fig.\ref{fig:delta_BG} at different times as function of $p_T$.
The deviations of Grad's approximation from actual distribution in BAMPS do not exceed 10\% for the
chosen value of $\eta/s$ below $p_T\sim 3 GeV$. The good agreement of BAMPS distribution with
Grad's approximation observed in Fig.\ref{fig:delta_BG} might indicate that the analytic expression
in Eq.\ref{foffeq_B} can be applied for derivations of hydrodynamic equations.

Derivation of the evolution equation for the shear stress tensor has been reported by us recently
in \cite{El10}. The starting point for the derivation is the entropy current which in kinetic theory
can be written as
\beq
s^\mu = -\int \frac{d^3 p}{E}  p^\mu f (\ln f - 1) \,.
\eeq
Writing the off-equilibrium distribution as $f=f_0\left( 1 + \phi \right)$ (comp. Eqs. \ref{foffeq}
and \ref{foffeq_B}) we expand the logarithm up to third order in $\phi$ and obtain by a direct
calculation
\beq
s^\mu \approx s_0 u^\mu - \frac{\beta_2}{2T} \pi_{\alpha\beta}\pi^{\alpha\beta} u^\mu -
\frac{8}{9}\frac{\beta_2^2}{T} \pi_{\alpha\beta}\pi^\alpha_\sigma \pi^{\beta\sigma} u^\mu
\label{smu}
\eeq
with $s_0 = 4n -n \ln \lambda$ and $\beta_2=\frac{9}{4e}$. Note that truncating the expansion at
order $\phi^2$ we obtain the entropy current from the Israel-Stewart theory \cite{IS,M04}.

The original Israel-Stewart equations explicitly satisfy the second law of thermodynamics since
they are obtained directly from the requirement of non-negativeness of the divergence of the entropy
current,  $\partial_\mu s^\mu \ge 0$. Using the same argumentation with a third-order entropy
current in Eq. \ref{smu}, i.e. calculating its divergence and imposing a linear relation between
the dissipative flux and the corresponding thermodynamic force, we obtain an extended version of
Israel-Stewart's equation for $\pi^{\mu\nu}$ in which we keep only terms up to third-order in
Knudsen number  $Kn\sim \tau_\pi \partial_\mu u^\mu$ resp. dissipative fluxes \cite{El10}:
\beq
\dot{\pi}^{\alpha\beta}  =  -\frac{\pi^{\alpha\beta}}{\tau_\pi} +
\frac{\sigma^{\alpha\beta}}{\beta_2} 
- \pi^{\alpha\beta} \frac{T}{\beta_2}\partial_\mu \left( \frac{\beta_2}{2T}u^\mu \right)
 -  \frac{8}{9} \frac{T}{\beta_2} \partial_\mu \left( \frac{\beta_2^2}{T} u^\mu
\right)\pi^{\langle \alpha}_{\sigma}\pi^{\sigma\beta\rangle}
\label{3rdOpi}
\eeq 
The latter equation constitutes a novel third-order evolution equation for the shear tensor. The
notation $\dot\pi$ denotes derivative with respect to the proper time $\tau$. $\tau_\pi$ is
the relaxation time and is the same as in Israel-Stewart theory:
\beq
\tau_\pi= 2 \eta \beta_2 .\
\eeq
Neglecting the last two terms in Eq. \ref{3rdOpi}
the second-order Israel-Stewart equation is recovered. To second order the equation we obtain does
not contain all terms found in recent works \cite{Romatschke:2009im,Rom10,Rom09,R09}. It is an
interesting task for the future to understand the differences between various formulations of
dissipative hydrodynamic equations, as discussed for example in \cite{R09}. We have to stress that
in our derivations, as presented here and in more detail in \cite{El10}, we do not use the Boltzmann
Equation directly. The terms we obtain are the full set of terms which can be obtained from the
entropy
principle if Grad's approximation, given by Eq. \ref{foffeq_B}, is used.

For a one-dimensional system with boost-invariance Eq. \ref{foffeq_B} takes the following form
\beq
\dot\pi=-\frac{\pi}{\tau_\pi}-\frac{4}{3}\frac{\pi}{\tau}+\frac{8}{27}\frac{e}{\tau}-3\frac{\pi^2}{
e\tau} .\
\label{3rdOpi1D}
\eeq
In the latter equation $\pi$ denotes the shear pressure which reduces the longitudinal pressure
$p_{L}$:
\beq
p_{L} = T_{33} = p - \pi .\
\eeq
Again, the Israel-Stewart equation is recovered from Eq. \ref{3rdOpi1D} if the last term is
neglected. Eq. \ref{3rdOpi1D} has to be solved together with the evolution equations for the energy
and particle densities. The former is obtained from the conservation of the energy-momentum tensor
component, $\partial_\nu T^{\nu 0} = 0$. The latter we obtain assuming conservation of the particle
number current, $\partial_\mu N^\mu = 0$, i.e. assuming a medium in which net particle number is
constant. In the geometry chosen here, the evolution equation for the energy and particle  densities
read 
\beq
\dot e=-\frac{4}{3}\frac{e}{\tau}+\frac{\pi}{\tau} ~~,~~ \dot n=-\frac{n}{\tau}.
\label{e} 
\eeq

Before discussing the solutions of Eq. \ref{3rdOpi1D} resp. of the Israel-Stewart's equations we
would like to discuss the effect of higher than 3rd order contributions to Eq. \ref{3rdOpi1D}. This
analysis will be presented here for a one-dimensional system. In order to include all
orders of corrections into Eq. \ref{3rdOpi1D} we assume they all have the form $x_n
\left( \frac{\pi}{
e}\right)^n\frac{e}{\tau}$ with $n \ge 3$. Note that the second and third-order terms in Eq.
\ref{3rdOpi1D} are already of this form. Thus, an \textit{ansatz} for an equation containing all
orders of corrections can be written in the following way
\begin{eqnarray}
\dot\pi &=& -\frac{\pi}{\tau_\pi}-\frac{4}{3}\frac{\pi}{\tau}+\frac{8}{27}\frac{e}{\tau}
+\sum_2^\infty x_n \left(\frac{\pi}{e}\right)^n\frac{e}{\tau} =  \nonumber \\
&=& -\frac{\pi}{\tau_\pi}-\frac{4}{3}\frac{\pi}{\tau}+\frac{8}{27}\frac{e}{\tau}
+\frac{\pi^2}{e\tau}\underbrace{\sum_2^\infty x_n \left(\frac{\pi}{e}\right)^{n-2}}_{\chi} =
\nonumber \\
&=& -\frac{\pi}{\tau_\pi}-\frac{4}{3}\frac{\pi}{\tau}+\frac{8}{27}\frac{e}{\tau}
+\chi \frac{\pi^2}{e\tau}   \label{all_O} .\
\end{eqnarray}
The coefficient $\chi$ is supposed to be an unknown function of time. Since the equation we consider
is supposed to include all orders of corrections, it should be applicable in the free-streaming,
i.e.  $\tau_\pi\to\infty$, limit. In the free-streaming limit the solutions for $e$,
$\pi$ and $n$ are known. Since the gas is streaming free, longitudinal pressure cannot be built up,
i.e. $p_L=0$, which means $\pi = p = \frac{e}{3}$. Using this, one finds that the energy density
evolves according to Eq. \ref{e} as $e(\tau) = e_0 \tau_0/\tau$. Using these solutions in the
$\tau_\pi\to\infty$ limit
in Eq. \ref{all_O} one obtains 
\beq
\chi = - \frac{5}{3} .\
\eeq  
This value is the result of resummation of higher-order terms in the heuristic \textit{ansatz} Eq.
\ref{all_O}. The equation obtained this way includes all orders of corrections, but only
approximately. If only third-order terms are included, one obtains $\chi = 3$, which corresponds to
Eq. \ref{3rdOpi1D}. The Israel-Stewart equation is obtained by setting $\chi = 0$. 

In the following we present the solutions of hydrodynamic equation of second (Eq. \ref{all_O} with
$\chi=0$) and third orders (Eq. \ref{all_O} with
$\chi=-3$)  and of the approximation of all orders (Eq. \ref{all_O} with
$\chi=-\frac{5}{3}$). The observable we use to quantify the deviations from equilibrium is the
pressure isotropy $\frac{p_L}{p_T} = \frac{p-\pi}{p+\pi/2}$. The results are compared to
BAMPS calculations using thermal initial conditions with $T_0 = 0.5 GeV$ and $\tau_0 = 0.4 fm/c$.
For the comparisons presented here only elastic processes with isotropic cross section are
included in BAMPS. The cross section $\sigma_{22}$ is parametrized to keep the $\eta/s$ value
constant, as has been already done in \cite{HM09, B09, El10}:
\beq
\sigma_{22}=\frac{6}{5}\left(\frac{\eta}{s}\right)^{-1}\frac{T}{4n-n\ln\lambda} .\
\label{cs22} 
\eeq   
where $4n-n\ln\lambda = s$. The method of derivation of hydrodynamic equations which has been
presented here does not allow to obtain an analytic expression for the shear viscosity coefficient.
We thus rely on the expression for $\eta$ obtained in \cite{DeGroot} which is strictly speaking
valid for the Israel-Stewart, i.e. second-order, theory.

\begin{figure}
\includegraphics[width=9.0cm]{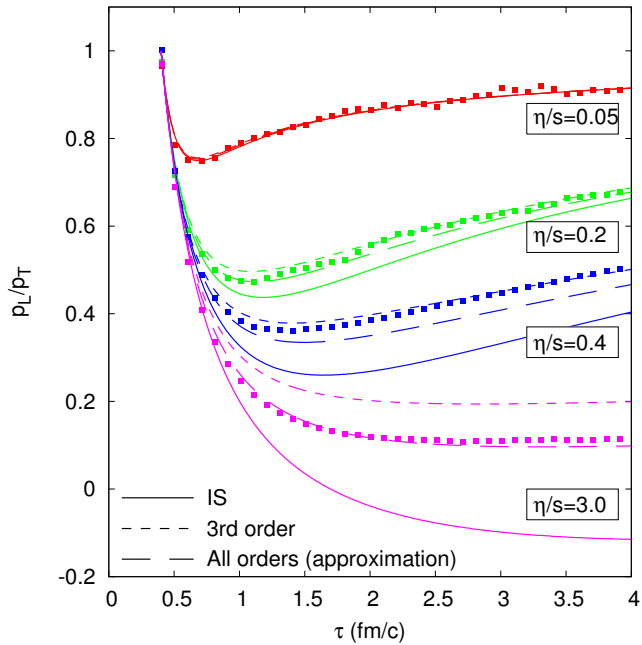}
\caption{Time evolution of pressure isotropy. The shear pressure is calculated by Eqs. \ref{e} and
\ref{all_O} using $\chi=0$ (Israel-Stewart), $\chi=-3$ (extension of Israel-Stewart's equation to
third order) and $\chi= - \frac{5}{3}$ (approximate inclusion of all order of corrections).}
\label{fig:piso}
\end{figure}
Time evolution of the pressure isotropy is presented in Fig.\ref{fig:piso}.  It was demonstrated
in \cite{B09} that the evolution of the system is governed by the Knudsen number
$Kn\equiv\tau_\pi/\tau \approx 6 \frac{1}{T\tau}\frac{\eta}{s}$. Since $T\approx T_0
\left(\tau_0/\tau \right)^{\frac{1}{3}}$, the Knudsen number depends in the situation considered
here on $\tau_0, \tau$, $T$ and $\eta/s$. Since all results are
obtained using the same initial conditions, the evolution depends only on the chosen value of
$\eta/s$, which is thus the direct measure of the strength of dissipative effects. The deviations
between the Israel-Stewart and BAMPS results become considerable already at $\eta/s=0.4$. At
$\eta/s=3$ the Israel-Stewart's equations ($\chi=0$ in Eq. \ref{all_O}) lead to a negative
longitudinal pressure which is not physical for the considered setup (This phenomenon has been as
well   investigated in \cite{Mart09}). The absolute deviations are reduced if one extends
Israel-Stewart's equations to third-order, Eq. \ref{3rdOpi1D} resp. $\chi=-3$ in Eq. \ref{all_O}.
Moreover, the negative longitudinal pressure does not occur in the solutions of the third-order
equations. Especially at late times, where the relaxation towards equilibrium sets in, the
third-order results on pressure isotropy are in very good agreement with BAMPS solutions. At early
times the third-oder and BAMPS results still deviate since there the expansion scalar $\partial_\mu
u^\mu=1/\tau$, is still large and thus all orders of corrections have to be taken into account.
Indeed, at early times solution of the approximate all-order equation, $\chi=-\frac{5}{3}$ in Eq.
\ref{all_O}, is in very good agreement with BAMPS results. This is due to the fact that the value
$\chi=-\frac{5}{3}$ has been obtained from the $\tau_\pi\to\infty$ or alternatively $Kn\to\infty$
limit, to which the system is close at early times.

In this study we have presented an extension of the Israel and Stewart's entropy based approach to
third order in dissipative fluxes. For this study we have considered a one-dimensional
boost-invariant gas of massless Boltzmann particles (gluons). Results of hydrodynamic
calculations have been compared to solutions of the Boltzmann Equation from the partonic cascade
BAMPS. Although the Grad's approximation, upon which the derivation of Israel-Stewart's equation is
based, has been shown to describe transverse spectra from BAMPS with remarkable accuracy, the
solutions of Israel-Stewart equations demonstrate large deviations from BAMPS. Inclusion of
third-order terms into the evolution equation for shear tensor reduces the deviations between
hydrodynamic and BAMPS results considerably. Up to $\eta/s=0.4$ the third-order solutions are in
very good agreement with BAMPS. In order to estimate the effect of all orders of correction, we have
introduced a resummation scheme which allows to represent the infinite series of higher-order terms
by one single term. The solution of this equation is in very good agreement with BAMPS results even
at $\eta/s=3$, when the Knudsen number is very large, which underlines the importance of
higher-order corrections at early times of evolution. It is still an important task for future
studies to understand the differences between the equations obtained here from the
entropy principle and the second-order equations presented in recent publications by different
authors \cite{Rom09, Romatschke:2009im, Denicol10, R09} . It is as well important to
investigate whether a possible extension of Grad's approximation can further improve the agreement
between hydrodynamic and kinetic transport calculations.

\section*{Acknowledgements}
The authors thank D.~H.~Rischke and P.~Huovinen for fruitful discussions, 
comments and their interest in this work. A.~E. thanks A.~Muronga for discussions and acknowledges 
the hospitality of UCT, Cape Town, where part of this work was accomplished and acknowledges support
by the Helmholtz foundation.
This work was supported by the Helmholtz International Center 
for FAIR within the framework of the LOEWE program launched by the State of Hesse.

%\section*{References}

%\section*{References}
%\bibliography{hq}

\end{document}